\documentclass[aps,prb,twocolumn,showpacs,amsmath,amssymb]{revtex4}
\usepackage{graphicx}
\usepackage{amsfonts}
\usepackage{amsmath, amsthm, amssymb}
\usepackage{dsfont}
\usepackage{mathptm}
\usepackage{color}
\usepackage{hyperref}
\usepackage{tabularx}
\definecolor{Red}{rgb}{1,0,0}

\newcommand{\be}{\begin{equation}}
\newcommand{\ee}{\end{equation}}

\newcommand{\ve}[1]{\mathbf{#1}}

\newcommand{\vd}{\ve{d}} 
\newcommand{\vg}{\ve{g}} 
\newcommand{\vk}{\ve{k}} 

\def\i{\mathrm{i}}

\begin{document}
\title[Proximity effect with noncentrosymmetric superconductors]{Proximity effect with noncentrosymmetric superconductors}

%
%

\author{Gaetano Annunziata}
\affiliation{Max-Planck-Institut f\"{u}r Festk\"{o}rperforschung, Heisenbergstrasse 1, D-70569 Stuttgart, Germany}

\author{Jacob Linder}
\affiliation{Department of Physics, Norwegian University of Science and Technology, N-7491 Trondheim, Norway}

\author{Dirk Manske}
\affiliation{Max-Planck-Institut f\"{u}r Festk\"{o}rperforschung, Heisenbergstrasse 1, D-70569 Stuttgart, Germany}

\date{Received \today}

\begin{abstract}
We describe the superconducting proximity effect taking place in a contact between a noncentrosymmetric superconductor and a diffusive normal/ferromagnetic metal within the quasiclassical theory of superconductivity. By solving numerically the Usadel equation with boundary conditions valid for arbitrary interface transparency, we show that the analysis of the proximity-modified local density of states in the normal side can be used to obtain information about the exotic superconductivity of noncentrosymmetric materials. We point out the signatures of triplet pairing, the coexistence of triplet and singlet pairing, and particular orbital symmetries of the pair potential. Exploiting the directional dependence of the spin polarization pair
breaking effect on the triplet correlations, we show how the order relation between triplet and singlet gaps can be discriminated and that an estimation of the specific gap ratio is possible in some cases.

\end{abstract}

\maketitle

\section{INTRODUCTION} \label{sec:int}
Unconventional superconductors are characterized by a pairing state breaking not only the $U(1)$ gauge symmetry. The most famous class of such materials is high-temperature cuprates in which anisotropic $d$-wave pairing is likely realized.~\cite{cupratesd} It is also known that electrons forming Cooper pairs need not to be bound in a spin-singlet state: Spin-triplet states with total spin quantum number $S=1$ are also possible.~\cite{triplet}
Several materials have been discovered showing strong signatures of a $p$-wave spin-triplet pairing state.~\cite{triplet_materials} A full understanding of pairing states in unconventional superconductors is still lacking but it is widely accepted that their origin cannot be explained within the Bardeen-Cooper-Schrieffer theory of conventional $s$-wave superconductors.~\cite{bcs} The determination of the symmetry of the superconducting order parameter is extremely important since it is often the first step to understand the mechanisms generating the superconducting phase.

Among unconventional superconductors a special place is occupied by noncentrosymmetric superconductors (NCSs), an intriguing class of compounds which has recently generated much attention due to its unique properties.~\cite{book} In these materials the superconducting phase develops in a low-symmetry environment with a missing inversion center. This broken symmetry generates a Rashba-type~\cite{rashba} spin-orbit coupling (SOC) and prevents the usual even/odd classification of Cooper pairs according to orbital parity, allowing a mixed-parity superconducting state. Since according to the Pauli principle even (odd) orbital parity states are associated with spin-singlet (triplet) states, disregarding for the moment the possibility of odd-frequency pairing,~\cite{odd} singlet and triplet Cooper pairs are allowed to coexist in NCSs.

The interest in these materials does not only stem from the lacking of a clear understanding of pairing-state symmetry and pairing mechanism. Indeed it has been recently recognized that NCSs can possibly be in a topologically nontrivial phase.
 This phase is believed to be characterized by line nodes, nontrivial topology of Bogoliubov-quasiparticles wavefunctions, zero-energy Majorana modes in vortex cores, and gapless edge states, all being stable and topologically protected against Hamiltonian parameters small variations.~\cite{topncs,yada11,schnyder12,brydon11,tanaka10} Moreover they are predicted to be suitable materials to generate and host spin polarized supercurrents.~\cite{vorontsov08,tanaka09,spinncs}

Since the discovery of noncentrosymmetric CePt$_3$Si, exhibiting a superconducting phase below T$_c=0.75 $K at ambient pressure,~\cite{bauer04} the family of NCSs has largely grown in number. Relevant examples are CeRhSi$_3$,~\cite{kimura05} CeIrSi$_3$,~\cite{sugitani06} CeCoGe$_3$,~\cite{settai07b} CeIrGe$_3$,~\cite{honda10} UIr,~\cite{akazawa04} Li$_2$Pd$_x$Pt$_{3-x}$B,~\cite{togano04} Mo$_3$Al$_2$C,~\cite{mo} and Y$_2$C$_3$.~\cite{yc} It is worth noting that superconductivity developing at oxide interfaces such as LaAlO$_3$/SrTiO$_3$ is intrinsically noncentrosymmetric.~\cite{reyren07} We will focus on the cerium compounds in what follows since within our formalism they will admit a common description. They share the same generating point group $C_{4v}$ lacking a mirror plane normal to the $c$ axis and have a qualitatively similar $T-P$ phase diagram, with a superconducting dome partially intersecting an antiferromagnetic phase.~\cite{ceptsi_pd,cerhsi_pd,ceirsi_pd,cecoge_pd,honda10}
They differ in the amplitudes and patterns of the ordered magnetic moments, they have a different electronic specific heat Sommerfeld coefficient and  effective masses. CePt$_3$Si is the compound with the largest Sommerfeld coefficient, a signature of strong electronic correlations,~\cite{fujimoto07,yanase} and it is the only one to become superconducting at ambient pressure. A detailed collection of available data on these material can be found in Ref.~\onlinecite{book} and references therein. The determination of superconducting pairing state effectively realized is still an open issue. Several experiments point toward an unconventional state with the possibility of lines nodes.~\cite{yogi04,bauer07,settai07a,yasuda04,izawa05bonalde05} The interpretation of experimental results can nevertheless be controversial since strong electronic correlations and SOC have been shown~\cite{fujimoto07,yanase} to influence the experimental findings.~\cite{yogi06} Indeed a triplet pairing state seems to be suggested by a critical magnetic field $H_{c2}$ larger than the condensation energy, e.g. beyond the Pauli-Clogston limit.~\cite{clogston} There is still controversy about this point since a superconducting phase developing close to an antiferromagnetic one would naturally correspond to singlet pairing state. In this scenario the large critical magnetic field could be explained taking into account the SOC.~\cite{frigeri04}
The standard route to identify a triplet pairing state is to exploit the interplay of superconductivity and magnetism, namely the magnetic field direction dependence of its pair breaking effect on Cooper pairs. While the pair breaking exist for any direction of the field for singlet superconductors, for triplet superconductors if the field is oriented parallel to the triplet equal spin pairing direction its effect is null. So measuring the spin susceptibility across superconducting transition can help identify triplet pairing states. However, the same experimental techniques used to ascertain that the prototypical triplet superconductor SrRuO$_4$~\cite{srruo}
has an equal spin pairing state with spin oriented in the $ab$ plane~\cite{ishida98} fail when applied to CePt$_3$Si.~\cite{yogi06}
While theoretical calculations including SOC predict a zero-temperature susceptibility value which is half the value of the one in the normal state when the field is in a pair-breaking configuration,~\cite{frigeri_samokhin} experiments~\cite{yogi06} find a temperature-independent susceptibility for any field direction. The impossibility to analyze the triplet pairing states with the common techniques is a consequence of the strong electronic correlations which enlarge the almost temperature-independent van Vleck term over the Pauli term in the spin susceptibility.~\cite{fujimoto07,yanase}
Another complication is that a magnetic field shifts the system in the mixed phase which in NCSs is predicted to be a helical vortex phase.~\cite{helical} We believe that these complications can be overcome by separating in space superconductivity and magnetism, and that an analysis of pairing states realized in NCSs can be put forth by considering hybrid contacts with ferromagnets. In particular analyzing the superconducting proximity effect~\cite{proxi} in such hybrids, the superconducting correlations propagating in the proximate region will be specific of the pairing state in NCSs but will not suffer from the complications described above.

The strategy of using hybrid contacts, junctions, and tunneling spectroscopy~\cite{wolf} to probe the pairing state of superconductors has been very fruitful. These techniques can be exploited to determine the gap amplitude in conventional superconductors and even the phase change or the existence of nodes in pairing states of unconventional superconductors.~\cite{tanakad,deutscher05} Many theoretical studies exist on tunnel and Josephson junctions with NCSs.~\cite{schnyder12,tanaka10,tanaka09,iniotakis07,linder07,wu09,yokoyama05,borkje06,asano11,sigristjos} Here we address the problem
of proximity effect in hybrid contacts of NCSs with normal and ferromagnetic metals. The most direct way to probe the proximity effect in experiments is to measure the local density of states (LDOS) in the nonsuperconducting region and to look at how it deviates as a function of position and energy from the normal-state LDOS for the influence of superconducting correlations. This can be done by placing contacts at different points of the normal region~\cite{gueron96} or with a STM tip capable of sweeping it.~\cite{sueur08} Our aim is to show that this kind of measurement can be useful to determine both the orbital and spin pairing state of NCSs. In particular we point out the signatures of triplet pairing, triplet and singlet pairing coexistence, and particular orbital pair potentials realizations. We show how the order relation between triplet and singlet gaps can be discriminated and that an estimation of their specific gap ratio is possible in some cases.

The remainder of this paper is organized as follows. In Sec.~\ref{sec:mod} the system is described and the formalism and equations employed are introduced. In Sec.~\ref{sec:res} we report our results for the proximity effect both in normal metals and ferromagnets. We compare different types of NCSs with different types of gaps. We also consider ferromagnets with varying exchange field amplitude and direction. Our conclusions are given in Sec.~\ref{sec:con}.

\begin{figure}[th!] \centering \resizebox{0.45\textwidth}{!}{
\includegraphics[width=80mm]{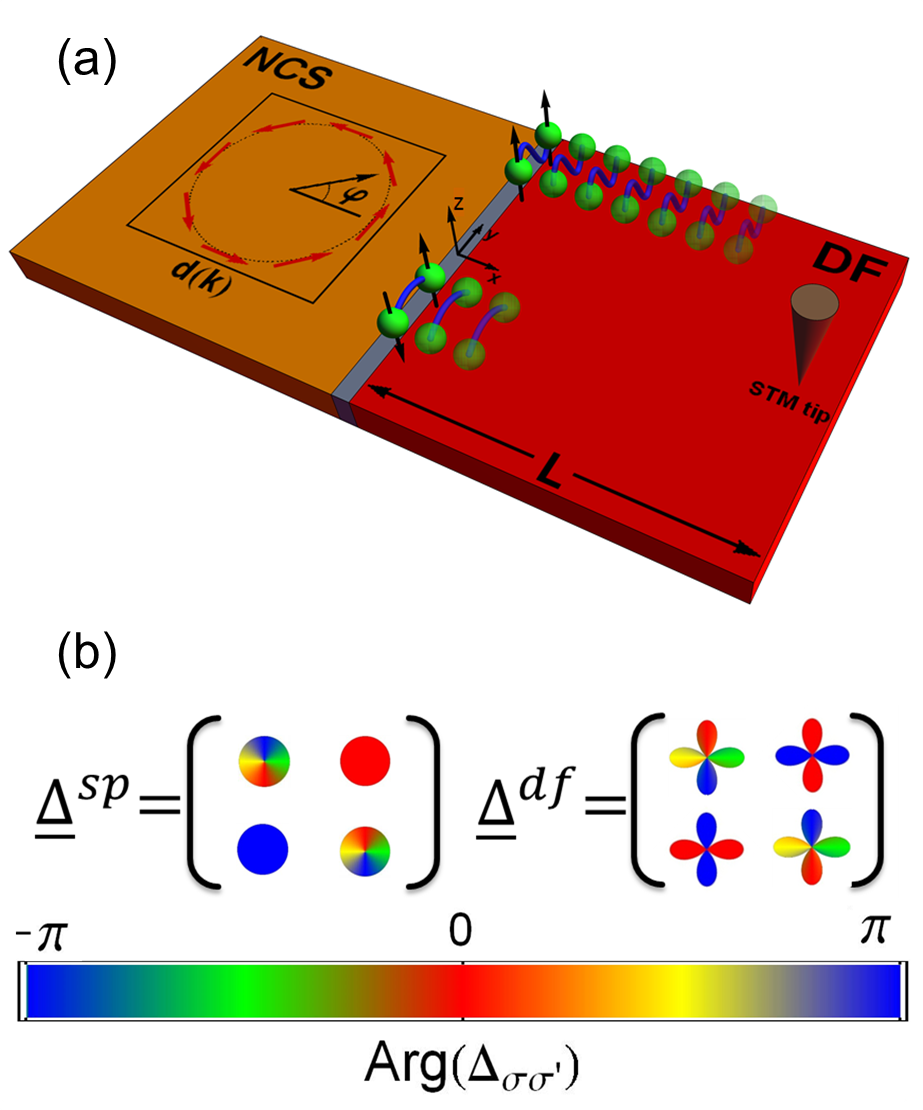}}
\caption{(Color online) In (a) the system under consideration is sketched. It is a contact between a NCS and a diffusive ferromagnet (DF) separated by an infinitely thin insulator. In such a setup the proximity effect features can be probed by looking at how the normal LDOS in DF is modified by the presence of the superconductor with, e.g., a position and energy resolved STM measurement. In the particular case of NCS both even-frequency spin-singlet and odd-frequency spin-triplet superconducting correlations are expected to propagate in the normal part of the contact; the latter can be long ranged depending on the relative orientation of exchange field in DF and d-vector in NCS. In (b) the amplitude and phase of the pair potentials matrices in two different types of NCSs are considered. Numerical simulations suggest that in cerium family NCSs $df$ potentials could be stabilized rather than the more commonly assumed $sp$ ones.~\cite{yokoyama}}
\label{fig:systempp}
\end{figure}
\section{MODEL AND FORMALISM} \label{sec:mod}
We consider an effective 2D contact between a clean noncentrosymmetric superconductor and a diffusive ferromagnet (NCS/DF) separated by a thin insulating barrier (see Fig.~\ref{fig:systempp}). The contact is at equilibrium at $T=0$. The system lies in the $xy$ plane with $x$ being the direction orthogonal to the interface. The NCS side is considered as a reservoir while the DF side is assumed to have a finite width $L$. We describe the contact within the quasiclassical theory of superconductivity, a formalism widely used in the analysis of superconducting heterostructures (see Ref.~\onlinecite{belzig99} for a general introduction and Refs.~\onlinecite{buzdin05} for its application to superconducting hybrids). The approximations behind this model have revealed being able to grasp
the main low-energy experimental features of heterostructures
involving conventional and unconventional superconductors,
both for ballistic
and diffusive systems.
The key consideration of the theory is that whenever the Fermi energy is much larger than other energy scales in the system, only a small error is carried out by considering only physical processes taking place close to the Fermi energy. Within this formalism, all the physical information is coded in the quasiclassical Green's function defined as
\be \label{eq:gcgf}
\hat{g}(\ve{r},\vk_F,\varepsilon) = \frac{\i}{\pi}
\int^\infty_{-\infty} \text{d}\xi_\vk \hat{G}(\ve{r},\vk,\varepsilon),
\ee
where $\hat{G}$ is the position, momentum, and energy dependent Fourier-transformed Gor'kov Green's function in the Wigner representation and $\xi_\vk=\vk^2/2m$. We have chosen units such that $\hbar=c=1$ and we use $\hat{\ldots}$
to denote $4\times4$ matrices in spin $\otimes$ Nambu space and $\underline{\ldots}$ for $2\times2$ matrices in spin space. While the momentum amplitude dependence has been integrated out, the quasiclassical Green's function still depends on Fermi momentum direction, i.e., $\vk_F=\left(\cos\varphi,\sin\varphi,0  \right)$, where $\varphi$ is the azimuthal angle in the $xy$ plane (see Fig.~\ref{fig:systempp}).
From now on we omit the subscript in the momenta since it is clear that in the quasiclassical theory they are always fixed on the Fermi surface. In nonequilibrium conditions a larger number of Green's functions together with proper distribution functions are necessary. This theory, known as Keldysh formalism, is redundant in our case and the retarded or advanced $4\times4$ Green's functions matrices are sufficient. They have the general structure
\begin{align}\label{eq:gstructure}
\hat{g}^{R,A} &= \begin{pmatrix}
 \underline{g}^{R,A} & \underline{f}^{R,A}  \\
-\underline{\tilde{f}}^{R,A} &  -\underline{\tilde{g}}^{R,A} \\
\end{pmatrix},
\end{align}
where $\underline{g}^{R,A}$ and $\underline{f}^{R,A}$ are the normal and anomalous propagator blocks, the latter being non zero only when superconducting correlations exist.
Each block has spin structure
\begin{align}\label{eq:gspinstructure}
\underline{g}^{R,A} &= \begin{pmatrix}
g_{\upuparrows}^{R,A} & g_{\uparrow\downarrow}^{R,A}  \\
g_{\downarrow\uparrow}^{R,A} &  g_{\downdownarrows}^{R,A} \\
\end{pmatrix},
\end{align}
and the tilde operation is defined as
\be\label{eq:tilde}
\underline{\tilde{f}}^{R,A}(\ve{r},\vk_F,\varepsilon)=\underline{f}^{R,A}(\ve{r},-\vk_F,-\varepsilon)^*.
\ee
Moreover the matrices are normalized such that
\be\label{eq:norm}
\hat{g}^{R,A}\cdot\hat{g}^{R,A}=\hat{1}.
\ee
Since retarded and advanced matrices are not independent in what follows we will use only the retarded matrix dropping the suffix $R$.
Once the latter is known, the LDOS can be calculated as
\be \label{eq:ldos}
N(\varepsilon)/N_0 = \frac{1}{2}\text{Re}\left[ \text{Tr} \left(\underline{g}\right)\right],
\ee
where $N_0$ is the density of states in the absence of superconductivity. Measuring the LDOS in the nonsuperconducting side of superconducting hybrids  is the most direct way to probe the proximity effect so we will focus on this quantity in what follows. The quasiclassical Green's function can be calculated by solving the Eilenberger equation,~\cite{eilenberger} valid in systems with arbitrary impurity scattering, and the Usadel equation,~\cite{usadel} valid in diffusive systems with strong impurity scattering. Since we are considering a contact between a clean superconductor and a diffusive ferromagnet, we have to solve different equations in the $x\gtrless 0$ regions and their solutions have to be connected at $x=0$ through a proper boundary condition. Our choice to consider different impurity scattering regimes follows from the consideration that superconductivity in noncentrosymmetric materials (and more generally unconventional non $s-$wave superconductivity~\cite{anderson59}) is known to occur only in rather clean samples.~\cite{bauer04,mineev07}

In NCSs the absence of a center of inversion generates an electric field which, according to special relativity, in the rest frame of electrons is felt as a magnetic field proportional to their orbital momenta, resulting in an effective antisymmetric Rashba spin-orbit coupling $\bf{g}_\vk=-\bf{g}_{-\vk}$. The particular form of this coupling for a noncentrosymmetric material can be determined by symmetry considerations. Moreover the broken inversion symmetry prevents the classification of Cooper pairs' wave function orbital parity which is allowed to be an admixture of even and odd parts. As a consequence in NCSs an admixture of triplet and singlet Cooper pairs can exist. It is worth noting that the Pauli principle \emph{allows} rather than \emph{dictates} the existence of a triplet pair potential in NCSs and that ideas to identify it are highly desirable. When considering the proximity effect in a NCS contact, both singlet and triplet correlations can manifest in the non superconducting side. Since these become isotropic in a diffusive medium, the triplet correlations have to be odd in frequency in order
to respect the overall antisymmetry dictated by the Pauli principle. This exotic type of correlation can decay in a magnetic medium on a scale typical of a non magnetic medium, at odds with standard even-frequency singlet correlations which are strongly suppressed by an exchange field.~\cite{bergeret,annunziata11} In a NCS the gap matrix in spin space has the general form
\be\label{eq:gapmat1}
\underline{\Delta}_\vk = \Delta^s f^s(\vk) \i\underline{\sigma_y} + \Delta^t f^t(\vk)
\mathbf{d}_\vk\cdot\boldsymbol{\underline{\sigma}}\i\underline{\sigma_y},
\ee
where $\Delta^{(s,t)}$ is the singlet (triplet) component amplitude, $f^{(s,t)}(\vk)$ are structure factors, and the well-known d-vector $\vd_\vk$~\cite{balian63} is used to parametrize the triplet pair potentials. The usual notation is used for Pauli matrices with $\underline{\boldsymbol{\sigma}}$ the vector of them. It is convenient to normalize Eq.~\eqref{eq:gapmat1} introducing the total gap amplitude $\Delta_0 \equiv  \max\limits_{\vk}  \text{Tr} \left( \underline{\Delta}_\vk^\dagger \underline{\Delta}_\vk  \right) /2$ and the triplet/singlet gap amplitudes ratio $r \equiv \Delta^t/\Delta^s \in [0,\infty)$, resulting in
\be\label{eq:gapmat2}
\underline{\Delta}_\vk = \Delta_0 \left(  \frac{f^s(\vk)}{\sqrt{1+r^2}} \i\underline{\sigma_y} + \frac{r f^t(\vk)}{\sqrt{1+r^2}}
\mathbf{d}_\vk\cdot\boldsymbol{\underline{\sigma}}\i\underline{\sigma_y}   \right).
\ee
With the parametrization employed the relative weight of triplet and singlet gaps amplitudes $r$ can be tuned without affecting the total gap amplitude $\Delta_0=\sqrt{\Delta^{s \, 2}+\Delta^{t\, 2}}$ and the extreme cases of pure singlet and triplet superconductors can be recovered for $r = 0$ and $r\to \infty$, respectively. The value assumed by the ratio $r$ in NCSs is extremely important since Andreev bound states can develop at the surface if $r>1$ $(\Delta^t>\Delta^s)$, depending on the orbital symmetry of pair potentials. Indeed it has been recognized that NCSs manifest a rich and peculiar bound-state structure.~\cite{schnyder12,brydon11,tanaka10} So it would be highly desirable to estimate experimentally the value of $r$ or at least discriminate between the $r\gtrless 1$ regimes. Another important point is to probe the structure factors and orbital symmetries of pair potentials. Several theoretical studies have pointed out the possible singlet and triplet pair potentials that are supposed to be realized in NCSs.~\cite{yanase,yokoyama,yada09} Part of our subsequent analysis is aimed toward these points. Similar ideas have been developed analyzing Raman scattering, tunneling conductance (without proximity effect) and Josephson effect features.~\cite{klam09,schnyder12,tanaka10,tanaka09,iniotakis07,linder07,wu09,yokoyama05,borkje06,asano11,sigristjos} It is widely accepted that the spin structure of triplet Cooper pair condensate is such that $\vd_\vk \parallel \bf{g}_\vk$ since this condition maximizes the superconducting critical temperature.~\cite{frigeri04} The form of the spin-orbit coupling vector $\bf{g}_\vk$ depends on
crystallographic point-group symmetries of the material.~\cite{samokhin09} We focus here on the most common family of NCSs: the cerium compounds. Their generating point group is the tetragonal point group $C_{4v}$ corresponding to $\mathbf{g}_\vk \propto (-k_y,k_x,0)$.~\cite{samokhin09} The resulting triplet pair potential has finite $S=1,m=\pm1$ and null $S=1,m=0$ components. Moreover $\Delta_{\upuparrows}$ and $\Delta_{\downdownarrows}$ have equal magnitudes resulting in an unitary triplet state. They occupy the diagonal elements of the gap matrix while the singlet elements are the off-diagonal ones.

The $\vk$ dependence of the pair potentials in NCSs is a matter of debate. The most common assumption is that they are the minimal spherical harmonics consistent with their spin structure, i.e., $s-$wave for singlet and $p-$wave for triplet pair potentials. Several studies suggest that this scenario is realized in NCSs.~\cite{yanase}
 This $sp$ case is obtained in our formulation by choosing $f^s(\vk)=f^t(\vk)=1$. The associated gap matrix $\underline{\Delta}^{sp}$ is depicted in panel (b) of Fig.~\ref{fig:systempp}. The singlet off diagonal functions have opposite sign while the triplet diagonal ones have opposite chirality $p_x\pm i p_y$. Theoretical studies point out that higher spherical harmonics pair potentials are possible.~\cite{yokoyama} The singlet pair potential can be $d_{x^2-y^2}-$wave and the triplet can be $f-$wave. The gap matrix in the $df$ case $\underline{\Delta}^{df}$ is depicted in panel $(b)$ of Fig.~\ref{fig:systempp} and is obtained in our formulation by choosing $f^s(\vk)=f^t(\vk)=k_x^2-k_y^2$. Other possibilities have been explored in the literature such as $f^s(\vk)=f^t(\vk)=2 k_x k_y$ which is related to the LaAlO$_3$/SrTiO$_3$ heterointerface~\cite{tanaka10} and $f^s(\vk)=2 k_x k_y$, $f^t(\vk)=1$ which is not related to any material but has been considered for theoretical interest.~\cite{asano11} Part of the subsequent analysis is devoted to the comparison of the $sp$ and $df$ scenarios, the two most probable pair potential realization in NCSs.


Several studies of NCSs within the quasiclassical theory of superconductivity exist in the literature.~\cite{vorontsov08,sigristjos,annunziata11,hayashi06} One often employed simplification is to neglect the spin-orbit coupling splitting of the Fermi surface since the Fermi energy is always much larger than the spin-orbit coupling. It has been pointed out how this simplification gives results qualitatively similar to the ones obtained without it.~\cite{schnyder12,tanaka10} The application of quasiclassical theory to a system with SOC depends on the strength of this interaction with respect to Fermi energy: For strong SOC one has to calculate two quasiclassical propagators in the two bands diagonalizing the SOC while for weak SOC the band splitting can be neglected and SOC can be considered as a self-energy.~\cite{book} While SOC can appear a strong interaction with respect to the superconducing gap, it is certainly weak with respect to the Fermi energy in the materials under examination~\cite{samokhin04} so we will adopt the second strategy. Since we are considering a NCS/DF contact where the NCS is in the clean limit, in the region $x<0$ its asymptotic quasiclassical Green's function $\hat{g}_S$ can be obtained from the Eilenberger equation
\be\label{eq:eil}
[\varepsilon\hat{\rho}_3 -
\hat{\Sigma}_{\vk} + \hat{\Delta}_{\vk}, \hat{g}_S] = \hat{0},
\ee
where $\hat{\rho}_3= \rm{diag}(1,1,-1,-1)$ and
\begin{align}\label{eq:socmat}
\hat{\Sigma}_\vk &= \begin{pmatrix}
\vg_\vk\cdot \underline{\boldsymbol{\sigma}} & 0\\
0 & [\vg_{-\vk}\cdot \underline{\boldsymbol{\sigma}}]^\mathcal{T} \\
\end{pmatrix}.
\end{align}
After a proper rotation in spin space~\cite{hayashi06} Eq.~\ref{eq:eil} can be solved analytically. Its solution for generic ratios of triplet and singlet gap amplitudes $r$ and for generic structure factors reads
\begin{align}\label{eq:eilsol}
\hat{g}_{S} &= \frac{1}{2}
\left(\hat{g}_{+}+\hat{g}_{-} \right),
\end{align}
where
\begin{align}
\hat{g}_{\pm} &= \begin{pmatrix}
 N_{\pm} & \mp i e^{-i\varphi} N_{\pm} & \pm i  e^{-i\varphi}  A_\pm &  A_\pm  \\
\pm i e^{i\varphi} N_{\pm} &  N_{\pm} & - A_\pm & \pm i  e^{i\varphi}  A_\pm \\
\pm i e^{i\varphi} A_{\pm} &  A_{\pm} & - N_\pm & \pm i  e^{i\varphi}  N_\pm \\
-A_{\pm} & \pm i e^{-i\varphi} A_{\pm} & \mp i  e^{-i\varphi}  N_\pm &  -N_\pm  \\
\end{pmatrix},
\end{align}
with normal and anomalous elements
\begin{align}
N_{\pm} &= \frac{\varepsilon}{\sqrt{\varepsilon^2-\frac{\Delta_0^2 f_\pm ^2}{1+r^2}}}, \\
A_{\pm} &= \frac{\Delta_0 f_\pm}{\sqrt{(1+r^2)\varepsilon^2-\Delta_0^2 f_\pm ^2}},
\end{align}
where
\begin{align}
 f_{\pm} &= f^s(\vk)+f^t(\vk) r.
\end{align}

The calculated Green's function does not depend explicitly on SOC magnitude but its structure is dictated by its existence. This is a well-known result of the application of quasiclassical theory to NCSs.~\cite{hayashi06} However, the absence of SOC is compensated in our formalism by the parameter $r$, since the amplitude of SOC has been shown to be related to the triplet gap amplitude in extended Hubbard model studies.~\cite{yokoyama}

The quasiclassical Green's function $\hat{g}$ for $x>0$ in the DF can be obtained from the Usadel equation
\begin{align}\label{eq:usadel}
D\partial_x(\hat{g}\partial_x\hat{g}) + \i[(\varepsilon+\i\delta)\hat{\rho}_3 +
\text{diag}[\mathbf{h}\cdot\underline{\boldsymbol{\sigma}},
(\mathbf{h}\cdot\underline{\boldsymbol{\sigma}})^\mathcal{T}],
\hat{g}]=\hat{0},
\end{align}
where $D=v_F^2\tau/2$, $\tau$ being the relaxation time associated with elastic impurity scattering, $\delta$ is an energy associated with inelastic scattering, e.g., quasiparticle damping, ``diag" indicates a diagonal block matrix, and $\mathbf{h}$ is the exchange field vector in DF. The characteristic energy scale of the diffusive side is the Thouless energy $E_{\text{Th}}=D/L^2$. Equation~\eqref{eq:usadel} can be solved with two boundary conditions at $x=0$ and $x=L$. While at $x=L$ the simple Neumann condition
\be\label{eq:boundL}
\partial_x \hat{g}|_L = \hat{0}
\ee
can be applied, the situation at $x=0$, where the solution of the Usadel equation has to be connected with the asymptotic solution of the Eilenberger equation [Eq.~\eqref{eq:eilsol}], is far more complicated. For contacts with conventional superconductors this problem has been solved and a  boundary condition valid for arbitrary interface scattering between the sides of the contact has been derived by Nazarov.~\cite{naz} A simplified version valid in the tunneling case had been derived by Kuprianov and Lukichev.~\cite{kl} The case of magnetic interface has also been treated.~\cite{magbc} In the case of unconventional superconductors a serious complication arises since the Green's function in the superconductor has a $\vk$ (or $\varphi$) dependence which is absent in the diffusive normal side since the strong impurity scattering isotropizes the Green's function. This problem has been solved by Tanaka et al.~\cite{tanaka_bc} who have generalized the Nazarov boundary condition for unconventional superconductors, both in a singlet and a triplet state. An infinitely thin insulating barrier located at $x=0$ gives rise to a finite scattering potential $H  \delta(x)$ which lowers the transmission probability $T$ as
\be\label{eq:T}
T(\varphi) = \frac{4 \
\cos^2(\varphi)}{4 \ \cos^2(\varphi) + \ Z^2},
\ee
where $\varphi$ is the angle with respect to the normal to the interface for a given trajectory (see Fig.~\ref{fig:systempp}), and $Z=2 m H /
k$ is the dimensionless barrier strength. The barrier resistance $R_B$ is then defined as
\be\label{eq:R}
R_B = \frac{2\ R_0}{\int_{-\pi/2}^{\pi/2}d\varphi\
T(\varphi)\cos(\varphi)},
\ee
where $R_0=4 \pi^2/k^2 A$ is the contact Sharvin resistance, $A$ being the constriction area. Introducing $\Gamma=R_B/R_F$, with $R_F$ the nonsuperconducting side resistance, the boundary condition at the interface can be written as
\be\label{eq:bound0}
\Gamma \  L \  \hat{g}|_0  \partial_x  \hat{g}|_0= 2
\langle\left[\hat{g}|_0, \hat{B}(\varphi)\right]\rangle,
\ee
where $\langle\ldots\rangle$ represents a transmission-mediated angular average on a half-Fermi surface,
\be
\langle f(\varphi)\rangle = \frac{\int_{-\pi/2}^{\pi/2}d\varphi\ f(\varphi)\cos(\varphi)}{\int_{-\pi/2}^{\pi/2}d\varphi\
T(\varphi)\cos(\varphi)},
\ee
and $\hat{B}(\varphi)$ depends on the asymptotic Green's function in the NCS. Defining
\be
T'(\varphi) = T(\varphi)/\left[2-T(\varphi)+2\sqrt{1-T(\varphi)}\right],
\ee
and
\be
\hat{H}_\pm(\varphi)=\frac{1}{2}\left[\hat{g}_S (\varphi)\pm \ \hat{g}_S
(\pi-\varphi)\right],
\ee
$\hat{B}(\varphi)$ can be written as
\begin{align}\label{eq:B}
\hat{B}(\varphi) = \frac{ -T'\left(\hat{1}+\hat{H}_-^{-1}
\right)+T'^2\hat{g}|_0\hat{H}_-^{-1}\hat{H}_+}{-T'\left[\hat{g}|_0,\hat{H}_-^{-1}\right]+
\hat{H}_-^{-1}\hat{H}_+ -T'^2\hat{g}|_0\hat{H}_-^{-1}\hat{H}_+\hat{g}|_0
}.
\end{align}
In Eq.~\eqref{eq:B} the fraction between matrices is only symbolic and should be understood as a matrix product between the inverse of the denominator and the numerator. The Usadel equation [Eq.~\eqref{eq:usadel}] and the boundary conditions [Eqs.~\eqref{eq:boundL},~\eqref{eq:bound0}] form a boundary value problem which cannot be solved analytically.
In the following section we report results obtained by solving numerically the boundary value problem with a finite difference code that implements the three-stage Lobatto IIIa formula and an initial guess for the Green's function is adapted self-consistently toward the final solution.

\section{RESULTS} \label{sec:res}
In order to look at the features of the proximity effect in the NCS/DF contact, once the quasiclassical Green's function has been obtained we calculate the LDOS in the non superconducting side with Eq.~\eqref{eq:ldos} as a function of energy and position. Notice that our LDOS is normalized with the normal state one such that when $N(\varepsilon)/N_0 = 1$ there is no proximity effect. Our main target is to find signatures in the proximity effect that can give hints on the nature of superconductivity in NCSs--namely, if there is triplet superconductivity, what is the ratio of triplet and singlet gap amplitudes?--distinguishing between $sp$ and $df$ pair potential symmetries. We use $\Delta_0$ as energy unit and the superconducting coherence length $\xi$ as the length unit. The energy is measured from the Fermi energy. The imaginary energy $i \delta$ in the Usadel equation [Eq.~\eqref{eq:usadel}], besides describing finite lifetime effects which are always present in real systems, has the advantage of stabilizing the code and facilitating the achievement of convergence. It is set to $\delta=0.01 \Delta_0$ while other parameters will take several values.

The formalism developed in the previous section allows us to explore limiting cases. The asymptotic solution for the Green's function in NCSs [Eq.~\eqref{eq:eilsol}] describes pure singlet and pure triplet superconductors for $r=0$ and $r\to \infty$, respectively. Moreover by neglecting the exchange field in the Usadel equation, the quasiclassical Green's function in a diffusive normal metal (DN) can be computed. We start exploring this case before considering the effect of ferromagnetism on the proximity effect.

\subsection{Proximity effect in NCS/DN}
\begin{figure}[th!] \centering \resizebox{0.45\textwidth}{!}{
\includegraphics[width=80mm]{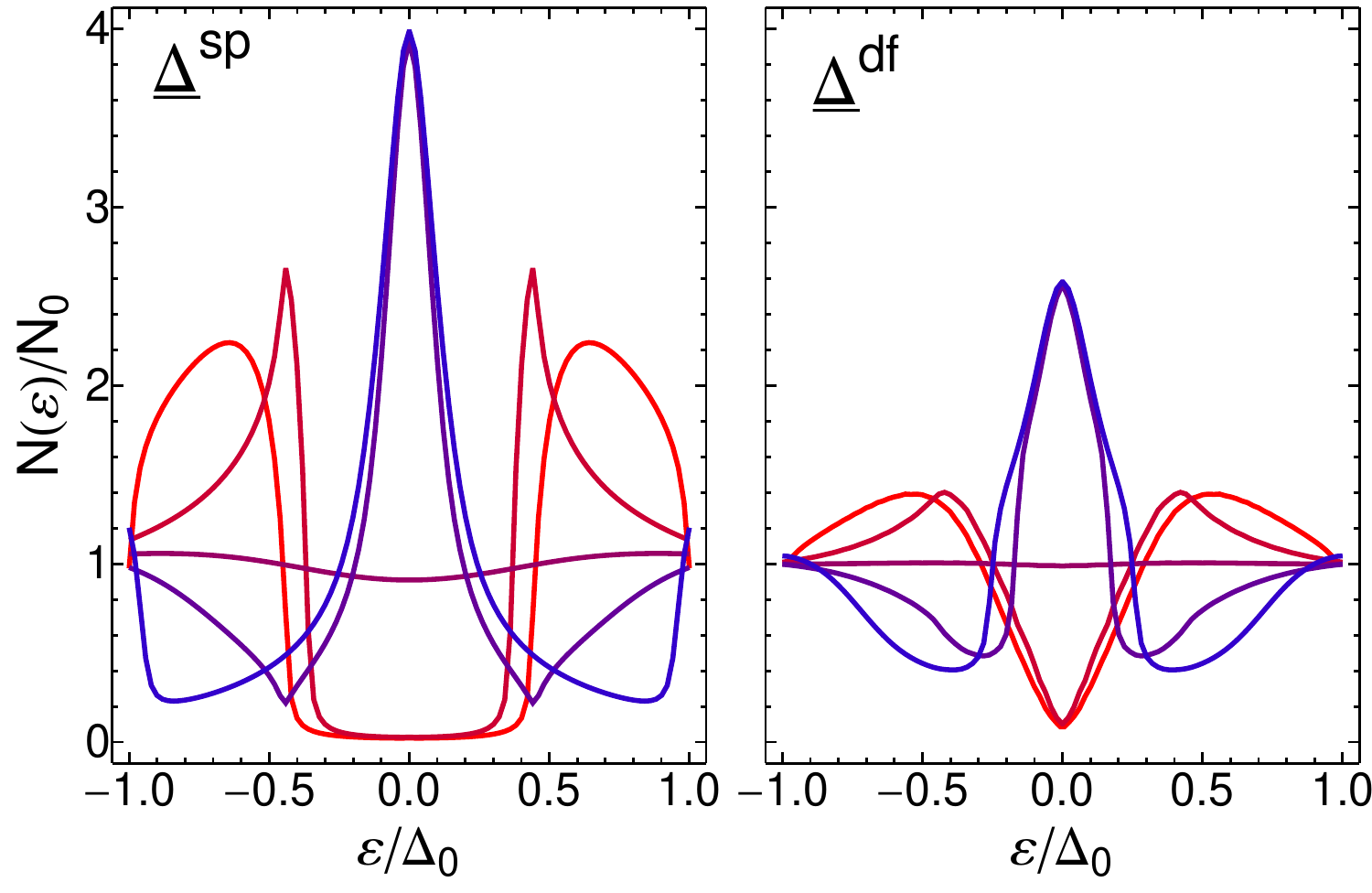}}
\caption{(Color online) LDOS calculated in the nonsuperconducting side
of a NCS/DN contact (see Fig.~\ref{fig:systempp}) for different NCSs.
Several $r=\Delta^t/\Delta^s$ values are plotted in both panels ($r=0,0.5,1,2,\infty$ from bottom to top at $\varepsilon=0$). $L=\xi$, $\Gamma=0.1$, $Z=2$ are fixed. The LDOS is calculated at $x=L$.}
\label{fig:NNCSvsE}
\end{figure}

In Fig.~\ref{fig:NNCSvsE} we show the LDOS as a function of energy comparing $sp$ and $df$ NCSs for several $r=\Delta^t/\Delta^s$ values ($r=0,0.5,1,2,\infty$ from bottom to top at $\varepsilon=0$ in both panels). The LDOS is calculated in the normal side at the border opposite to the interface, i.e., at $x=L$ (see Fig.~\ref{fig:systempp}), and its width is fixed to $L=\xi$. In a typical setup $L \sim 10$ nm. The interface between the NCS and DN is assumed to be in an intermediate transparency regime with $\Gamma=0.1$ and $Z=2$. A change in the values of these parameters modifies quantitatively the LDOS but not its general features, except for $L$ since the proximity effect eventually disappears farthest from the interface. In the $sp$ case for $r=0$, i.e., without a triplet gap, the typical proximity effect between an $s-$wave superconductor and a normal metal is recovered. Its signature is the opening of a minigap~\cite{millan68} in the LDOS of the normal metal. Its width depends on junction parameters such as interface transparency and normal layer length.~\cite{volkov95} In a NCS with $0<r<1$, e.g., $r=0.5$, the minigap is still there but its width is smaller with respect to the pure $s-$wave case. Assuming equal amplitude for triplet and singlet gaps, i.e., $r=1$, the LDOS is almost flattened at its normal-state value and a weak proximity effect exists only close to the Fermi energy $\varepsilon=0$. This peculiar behavior can be regarded as a signature of a transition to a topologically nontrivial phase.
Indeed the features of the LDOS change abruptly for $r>1$. The main feature is a well-defined zero-energy peak (ZEP) which is associated with the proximity effect from triplet superconductors.~\cite{tanaka_pxy} In the $df$ NCS case the LDOS is quantitatively less modified by the proximity effect since the pair potentials are not fully gapped. The $r$ dependence of the LDOS in this case is similar to the $sp$ case but the energy dependence in the $r\gtrless 1$ regimes differs. Indeed a dip rather than a minigap exists in the LDOS at low energy for $r<1$. This is the typical situation in a [100] contact between a high T$_c$ cuprate and a normal metal while a [110] interface, i.e., an effective $d_{xy}$ rather than a $d_{x^2-y^2}$ pair potential, would not generate any proximity effect.~\cite{ohashi96,tanaka0405,sharoni04} The dip becomes narrower increasing $r$ as long as $r<1$. Again there is almost no proximity effect for $r=1$ and a ZEP develops for $r>1$. This peak is different from the one in the $sp$ case since shoulders are present at finite energy giving it a bell shape. The shoulders are a signature of additional peaks at finite energies reflecting the nonmonotonic dispersion of bound states in the $df$ case.~\cite{tanaka10} This peculiar type of peak is clearly distinguishable from a standard ZEP in
tunneling experiments.~\cite{koren12} The LDOS features described depend quantitatively but not qualitatively on the contact parameters; i.e., the width of the minigaps and the width and height of the peaks depend on the particular values of $L$, $Z$, and $\Gamma$, but do not disappear as long as proximity effects take place. We have chosen an intermediate transparency interface even if this is not the usual experimental situation only for the sake of clarity. The main difference in considering a tunneling interface is that the peaks and minigaps appear more narrow, but their features and the possibility of their distinction remain unaltered. At the end of this section the more proper case of tunneling interface will be considered when we report on the penetration length of superconducting correlations in the normal side. From the analysis of the proximity effect in NCS/DN several considerations can be made. For both the $sp$ and $df$ cases the regimes $r\gtrless 1$ are clearly distinguishable by the low-energy LDOS, i.e., minigap for $r<1$ and ZEP for $r>1$ in the $sp$ case and dip for $r<1$ and ZEP for $r>1$. The same differences can be exploited to distinguish between $sp$ and $df$ NCSs. Even for $r>1$ where they both show a ZEP in the LDOS, a distinction is possible since only in the $df$ case the peak has shoulders at finite energies due to the nonmonotonic dispersion of the bound states. The determination of the particular value of $r$ rather than the regime $r\gtrless 1$ appears to be more difficult. Indeed it is clear that the features of the LDOS in NCS/DN are mostly dominated by the larger gap. In other words a NCS with singlet gap larger than the triplet one gives rise to a proximity effect very similar to the one from a pure singlet superconductor and vice versa. This is a general behavior of NCSs not only manifested in the proximity effect. However, there are quantitative dependencies on the particular value of $r$ assumed in both regimes $r\gtrless 1$ such as the width of the minigap and the height of the ZEP such that in principle a fitting procedure can be employed to determine its value from the experimental data.~\cite{gueron96} A simpler estimation is nevertheless possible by looking at the proximity effect in NCS/DF.

\begin{figure*}[th!] \centering \resizebox{1\textwidth}{!}{
\includegraphics[width=80mm]{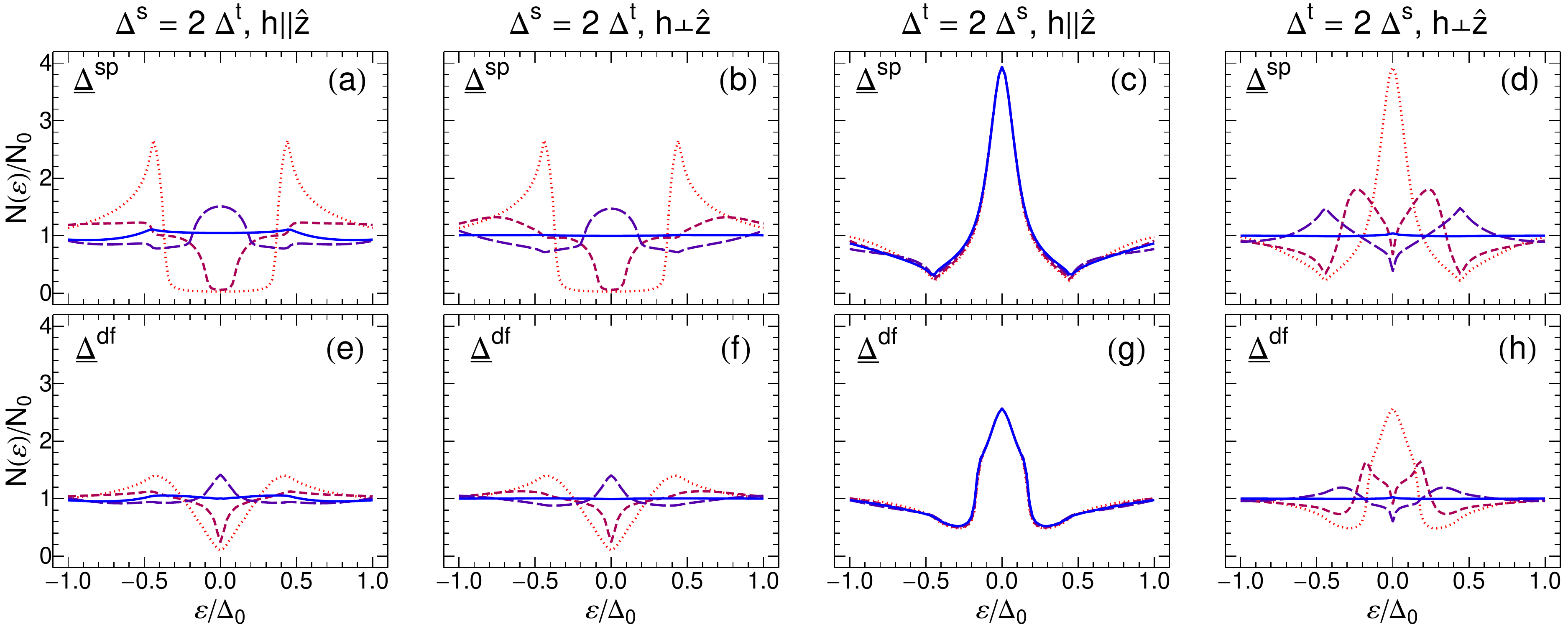}}
\caption{(Color online) LDOS calculated in the nonsuperconducting side
of a NCS/DF contact (see Fig.~\ref{fig:systempp}) for different NCSs, gaps ratio values, exchange field amplitudes and directions in DF.
The NCS/DN case (dotted lines) is plotted for comparison. Short-dashed, long-dashed, and solid lines are for $h/\Delta_0=0.5,1,5,$ respectively.
$L=\xi$, $\Gamma=0.1$, $Z=2$ are fixed. The LDOS is calculated at $x=L$.}
 \label{fig:FNCSvsE}
\end{figure*}

\subsection{Proximity effect in NCS/DF}
The analysis of the proximity effect in a NCS/DF junction can ease the identification of triplet pairing in NCSs, the distinction between different types of NCSs, and the estimation of the triplet/singlet gap ratio $r$. Even if the simpler NCS/DN analysis can give sufficient information in this sense, loopholes can exist. The point is that observing ZEPs in LDOS is not sufficient to claim the existence of a triplet pairing in the superconductor since the peak could be generated by unconventional singlet superconductivity or could indeed be a signature of triplet correlation but generated from spin-sensitive processes at the interface or magnetic impurities rather than by a real triplet pair potential in the superconductor. The idea is to exploit the pair-breaking effect of the exchange field on the Cooper pairs. While a spin-polarized environment tends always to destroy singlet Cooper pairs, this is not true for triplet Cooper pairs where a polarization parallel to the equal spin pairing direction has no effect at all.~\cite{bergeret,annunziata11,brydon} Considering that the d-vector of triplet superconductivity is built such that the equal spin pairing direction is in the plane orthogonal to it, this condition reads ${\bf h} \bot \mathbf{d}_\vk$. If ${\bf h}$ has some components parallel to $\mathbf{d}_\vk$ pair breaking takes place even for triplet Cooper pairs.
Indeed the standard route to identify triplet pairing is to measure the polarization direction dependence of spin susceptibility across the superconducting transition. While this technique has been proven successful for Sr$_2$RuO$_4$, the prototypical triplet superconductor, it fails when applied to NCSs due to strong electronic correlations. A possible way to overcome this difficulty is to extract the superconducting correlations by the complicated environment of the NCS and analyze how they react to a pair-breaking field in a much simpler environment such as a ferromagnet, that is, study the proximity effect in NCS/DF. In our case the d-vector lies in the $xy$ plane so an exchange field out of the plane of the junction ${\bf h}\parallel \hat{{\bf z}}$ is not pair breaking while any finite component in the $xy$ plane is pair breaking for triplet correlations. For singlet correlations the exchange field is pair breaking for any orientation. Figure~\ref{fig:FNCSvsE} shows the LDOS calculated in the nonsuperconducting side of a NCS/DF contact for different NCSs (the first row for the $sp$ type and the second for the $df$ type). The two panels of a column differ only in the type of NCS with all other parameters being equal. Both $r\gtrless 1$ regimes are shown for two possible orientations of the exchange field ${\bf h}\parallel \hat{{\bf z}}$ and ${\bf h}\bot \hat{{\bf z}}$. The NCS/DN case (dotted lines) is plotted for comparison and in each panel short-dashed, long-dashed, and solid lines differ in the amplitude of the exchange field, $h/\Delta_0=0.5,1,5,$ respectively. $L=\xi$, $\Gamma=0.1$, $Z=2$ are fixed and the LDOS is calculated at $x=L$.
Panels $(a), (b), (e), (f),$ show how the exchange field modifies the LDOS when the NCSs are mostly singlet superconductors, i.e., $\Delta^s=2\Delta^t$. In this case the minigap and the dip existing in the NCS/DN case are narrowed for low fields, while for fields of the order of the superconducting gap an enhancement of LDOS is possible and ZEPs in the $df$ case or domes in the $sp$ case can develop. These different shapes can be used to distinguish between the two types of NCSs. The enhancement rather than the suppression of LDOS is indeed found in measurement of contacts between singlet superconductors and ferromagnets.~\cite{kontos01} For larger exchange field values such as to overcome the Pauli-Clogston limit, the pair-breaking effect eventually becomes so strong as to destroy completely the proximity effect [see solid curves of panels (b) and (f)]. However in a situation of absence of pair breaking for the triplet correlations a residual weak proximity effect can exist [see solid curves of panels (a) and (e)]. This can be exploited as follows. Reporting the NCS/DN case we have concluded that a NCS with singlet gap larger than the triplet one gives rise to a proximity effect very similar to the one from a pure singlet superconductor and vice versa, and that a distinction between the extreme cases of pure triplet/singlet and mixed cases in the regimes $r\gtrless 1$ requires a quantitative rather than qualitative analysis. In NCS/DF the situation is different. The fact that the exchange field is pair breaking for every direction for singlet correlations but not for triplet correlations gives the possibility to distinguish between a pure singlet case (conventional or not) and a real mixed case where singlet and triplet gaps exist. Indeed the presence of a subdominant triplet gap can be ascertained by a residual proximity effect in strong ferromagnets with exchange field in a non-pair-breaking configuration for the triplet correlations [compare the solid lines of panels (a), (b), (e), and (f)]. Of course in order to observe the pair-breaking effect it is not necessary to have a field totally in plane but a finite in-plane component is enough. Panels (c), (d), (g), and (h) show how the exchange field modifies the LDOS when the NCSs are mostly triplet superconductors; i.e., $\Delta^t=2\Delta^s$. In this case when the field is not in a pair-breaking direction [see panels (c) and (g)], its effect on the LDOS is minimal. At low energies there is no difference from the NCS/DN case, while small quantitative deviations exist at larger energy. The reason for the insensitivity of the LDOS to the subdominant component in the $r>1$ regime lies in the fact that the NCS is in a topologically non trivial phase with Andreev bound states at the interface. These states totally dominate the physics of the contact. Once they have been triggered by the condition $r>1$, they become only weakly dependent on the magnitude of the subdominant singlet gaps rendering the LDOS almost insensitive to an exchange field which is pair breaking only for the singlet correlations.~\cite{asano11} In particular the zero-energy bound state with momentum orthogonal to the interface does not depend at all on the subdominant singlet component and only the bound states associated with large-angle trajectories have an appreciable dependence on it. Moreover the larger contribution to the proximity effect comes from almost orthogonal trajectories since they are associated with a larger transmission probability [see Eq.~\eqref{eq:T}]. When the exchange field is pair breaking for triplet correlations [see panels (d) and (h)], the peaks observed in the NCS/DN case split and lower until eventually the proximity effect disappears. The different behavior for in-plane and out-of-plane fields can be exploited to ascertain whether the peaks observed in the LDOS of NCS/DN are really generated by triplet correlations. Indeed only in this case a sensitivity on the exchange field direction in NCS/DF should be present.

\begin{figure}[th!] \centering \resizebox{0.45\textwidth}{!}{
\includegraphics[width=80mm]{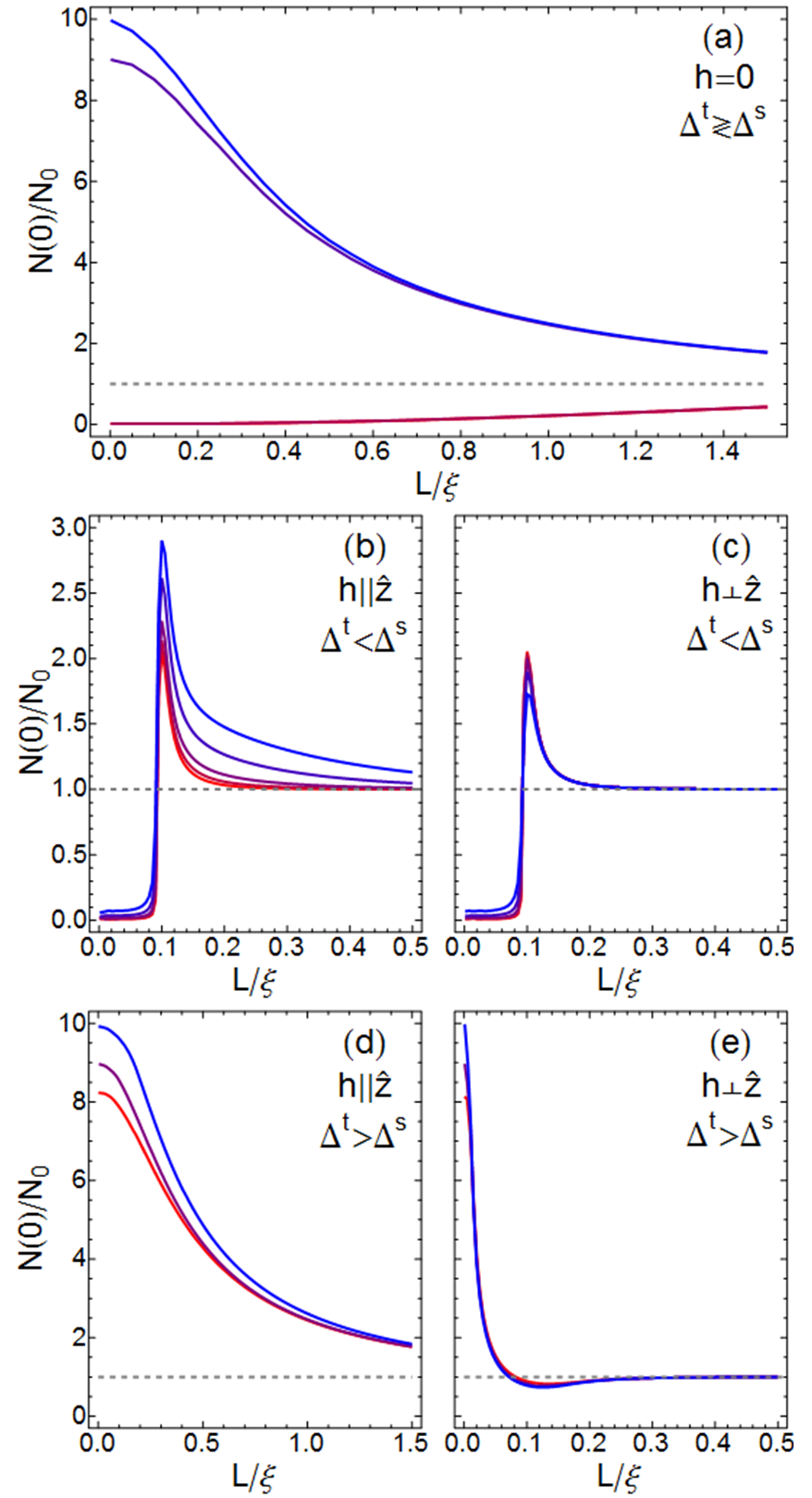}}
\caption{(Color online) Zero-energy LDOS calculated in the non superconducting side
of a NCS/DN contact (a) and of a NCS/DF contact [(b)--(e)] as a function of the normalized distance from the interface $L/\xi$. Here only the $sp$ symmetry is plotted since at zero energy the $df$ symmetry results differ only quantitatively. Everywhere $\Gamma=10$, $Z=10$ are fixed and a dashed line representing the absence of proximity effect is plotted. In (a) $r=\Delta^t/\Delta^s=0,0.5,2,\infty$ are considered from bottom to top (the first two are superposed and practically indistinguishable). In (b) and (c) the case $\Delta^t<\Delta^s$ is considered for two orientations of the exchange field in the non superconducting side and $h/\Delta_0=5$. The cases $r=0,0.4,0.6,0.8,0.9$ are plotted from bottom to top close to the interface. In (d) and (e) the case $\Delta^t>\Delta^s$ is considered for two orientations of the exchange field in the non superconducting side and $h/\Delta_0=5$. The cases $r=1.5,2,\infty$ are plotted from bottom to top close to the interface and everywhere the LDOS is calculated at $x=L$.}
 \label{fig:vsL}
\end{figure}

\subsection{Penetration length of superconducting correlations}
The spatial dependence of the LDOS is worth exploring since the penetration depth of superconducting correlations in the normal side is also a source of information about the pairing state. Figure~\ref{fig:vsL} shows the zero-energy LDOS in the nonsuperconducting side of the contact as a function of the normalized distance from the interface $L/\xi$. Only the $sp$ symmetry is plotted since at zero energy the $df$ symmetry results differ only quantitatively. In each panel $\Gamma=10$, $Z=10$ are fixed and a dashed line representing the absence of proximity effect is plotted. In panel (a)
the case of NCS/DN is shown for several $r$ values, $r=\Delta^t/\Delta^s=0,0.5,2,\infty$ from bottom to top (the first two curves are superposed). It is clear that the range of proximity effect in NCS/DN does not depend on the triplet/singlet gaps ratio, but when the triplet gap is dominant the magnitude of the proximity effect is much more sensitive to the distance from the interface. The range of the proximity effect can have an $r$ dependence in NCS/DF. Panels (b) and (c) are for a NCS/DF contact with a mostly singlet NCS and with an exchange field out of plane and in plane and $h/\Delta_0=5$. The cases $r=0,0.4,0.6,0.8,0.9$ are plotted from bottom to top close to the interface. The oscillation of LDOS around the normal-state value and the faster decay with respect to the nonmagnetic case typical of singlet superconductor/ferromagnet contacts are present for both orientations of the field. However when the field is not pair breaking on the subdominant triplet correlations an $r$ dependence of superconducting penetration length clearly appears. The larger the triplet component the larger the range of the proximity effect. Since the penetration length is easy to measure and the quasiclassical theory is able to give a good estimation for it from the contact parameters in a conventional superconductor/ferromagnet structure,~\cite{kontos01} the range of the proximity effect could be a precious source of information for NCS superconductivity since an unexpectedly long range proximity effect would be a strong signature of the existence of a subdominant triplet component. This could also be ascertained in the NCS's Josephson junction showing a coupling on distances larger than expected. When the NCS is mostly a triplet superconductor this sensitivity disappears as a consequence of Andreev bound state formation as explained before. Panels (d) and (e) shows this case with an exchange field out of plane and in plane and $h/\Delta_0=5$. The curves plotted are for $r=1.5,2,\infty$ bottom to top close to the interface. In this case the proximity effect is long ranged for out of plane field. Indeed it decays on the same scale of a NCS/DN contact [see panel (a)]. In a pair-breaking field configuration oscillations around the normal-state LDOS appear and the penetration length is reduced as in a NCS/DF contact with $r<1$ [see panel (c)]. The $r$ dependence in the $r>1$ regime is weak and only small quantitative differences emerge close to the interface. One could try to overcome this by analyzing the range of correlations at finite energies rather than zero energy; however the $r$ dependence that emerges is always small compared to the one in the $r<1$ regime. This could be seen by the small deviation induced by $h$ in panel (c) and (g) of Fig.~\ref{fig:FNCSvsE}. From the analysis of the proximity effect in NCS/DF several considerations can be made. The interplay of ferromagnetism and superconductivity can give extra information with respect to the NCS/DN case. The $sp$ and $df$ types can be distinguished; real equal spin pairing correlations generated by a triplet gap in NCS can be discriminated by directional dependence of the pair-breaking exchange field in both energy and space resolved analysis; in NCSs with a dominant singlet gap the existence of a subdominant triplet gap and an estimation of triplet/singlet gap ratios are possible.

\section{CONCLUSIONS} \label{sec:con}
We have studied the proximity effect in a 2D contact between a clean noncentrosymmetric superconductor and a diffusive metal and ferromagnet within the quasiclassical theory of superconductivity. To look at the features of the proximity effect we have analyzed how the presence of the noncentrosymmetric superconductor influences the LDOS in the nonsuperconducting side by numerically solving the boundary problem determined by the quasiclassical Usadel and Eilenberger equations. We have found unique signatures in the proximity effect of the exotic superconductivity expected in NCSs and our findings can be easily exploited in tunneling spectroscopy experiments, overcoming the difficulties encountered in spin susceptibility measurements.
Focusing on the most common family of noncentrosymmetric superconductors, the cerium compounds, we have found that the existence of a dominant triplet gap can be clearly ascertained by a zero-energy peak in the LDOS measured in a proximate normal metal. We have also shown that the features of the LDOS
are mostly dominated by the larger gap, a noncentrosymmetric superconductor with singlet gap larger than the triplet one giving
rise to a proximity effect very similar to the one from a pure
singlet superconductor and vice versa. However, quantitative differences exist and the particular value of the triplet/singlet gap specific ratio could be determined by a fitting procedure of experimental data. Unique signatures of orbital symmetries exist for both triplet and singlet dominating gaps. So the proximity effect can distinguish between $sp$ and $df$ orbital symmetries. For a dominant singlet gap the $sp$ symmetry induces a minigap and the $df$ a dip in the normal metal. For a dominant triplet gap both induce a zero-energy peak but in the $df$ case it is accompanied by weaker peaks at finite energies. The range of superconducting correlations penetration in the normal metal does not depend on the orbital symmetries and on the gap amplitude ratio. The analysis of proximity effect in a ferromagnet gives additional insights. We have shown that it can help to ascertain whether the peaks seen in the normal metal case are really a signature of a dominant triplet gap in the noncentrosymmetric superconductor and not a spurious effect generated by magnetic impurities or spin-sensitive processes at the
interface since a genuine triplet zero-energy peak is insensitive to an exchange field oriented out of the junction plane while it is destroyed by a sufficiently strong in-plane field. In order to
change the orientation of the field, two separate samples where the
exchange field is locked to different orientations via, e.g.,
antiferromagnetic coupling, or with different crystallographic orientations  can be grown to effectively change the
orientation of the exchange field. We have shown that another signature of triplet dominant gap is that the zero-energy peak survives far inside the non superconducting side despite the exchange field when it is in a non-pair-breaking configuration. The exchange field can be exploited again to distinguish between $sp$ and $df$ types of noncentrosymmetric superconductors since for a field of the order of the superconducting gap they show different low-energy behaviors, $sp$ showing a dome and $df$ a peak in the regime where the singlet gap is dominant. We have shown that in this regime the presence of a subdominant triplet component can be ascertained by the existence of a residual proximity
effect in strong ferromagnets with exchange field in a non-pair-breaking configuration for the triplet correlations. This effect can be easier to observe by looking at the penetration length of superconducting correlations in the ferromagnet since
the larger the subdominant triplet component the larger
the range of the proximity effect. Since this quantity is easy
to measure and the quasiclassical theory is able to give a good
estimation for it from the contact parameters in a conventional
superconductor/ferromagnet structure, the
range of the proximity effect could be a precious source of information
for NCS superconductivity since an unexpectedly
long range proximity effect would be a strong signature of
the existence of a subdominant triplet component. The specific value of $r$ can be deduced by the enhancement in the penetration length in this case. The identification of a subdominant singlet component in a mostly triplet noncentrosymmetric superconductor appears more subtle instead since the the Andreev bound states existing in this regime are
only weakly dependent on the magnitude of the subdominant
singlet gaps rendering the LDOS almost insensitive to an exchange
field pair breaking only for the singlet correlations.

The influence of interface scattering on the triplet pair potential of NCSs deserves a comment since it is expected to be pair breaking for anisotropic superconducting order parameters.~\cite{buchholtz81} This effect cannot be easily incorporated in our model since we are solving different equations for each side of the contact and they are connected by a boundary condition containing an integration. However, it has been shown~\cite{vorontsov08} that taking into account this effect the surface Andreev bound states do not disappear but merely change dispersion for large-angle trajectories. Since the contribution to the proximity effect of a trajectory is inversely proportional to its angle, that would introduce only small quantitative changes and does not affect our conclusions. With the ongoing activity of characterizing exotic superconducting
materials where triplet and mixed pairing are believed to be present we believe our results on proximity effect may serve as a useful tool to experimentally identify superconducting properties of NCSs.

\acknowledgments
We thank A. Schnyder, Y. Tanaka, P. M. R. Brydon, and G. Koren for discussions and clarifications. G.A. thanks the Department of Physics of Norwegian University of
Science and Technology for hospitality.

\end{document}